# Weather event severity prediction using buoy data and machine learning

Vikas Ramachandra
Adjunct Professor, Data Science
University of California, Berkeley
Berkeley, CA
virama@berkeley.edu

## Abstract

In this paper, we predict severity of extreme weather events (tropical storms, hurricanes, etc.) using buoy data time series variables such as wind speed and air temperature. The prediction/forecasting method is based on various forecasting and machine learning models. The following steps are used. Data sources for the buoys and weather events are identified, aggregated and merged. For missing data imputation, we use Kalman filters as well as splines for multivariate time series. Then, statistical tests are run to ascertain increasing trends in weather event severity. Next, we use machine learning to predict/forecast event severity using buoy variables, and report good accuracies for the models built.

## Introduction and background

NOAA provides detailed historic data for extreme weather events, particularly tropical storms, hurricanes, tornadoes and high wind gusts, as well as physical variable recordings from sensors at buoy stations, such as air temperatures, wind speeds, wace heights and so on [1][2]. This data in publicly available and is useful as input to data analyses. Previous work in this domain can be summarized as follows. In [6], the authors use Bayesian statistical techniques to analyse storm surges and combine them with physical models for better accuracy. Also, in [7], the authors use machine learning models including neural networks to predict wind-wave levels. More recently, in [8], the authors use interpretable convolutional neural networks to predict hail size in extreme hailstorms. To add to the above body of work, in this paper, we propose the following algorithmic pipeline: Spatio-temporal missing data imputation, followed by trend analysis and machine learning--Xgboost based time series forecasting for high wind based weather events such as tropical storms and hurricanes.

## Data sources

We analyzed data for a number of buoy stations manually picked from the NOAA interactive map (figure below), as well as storm event data (intensity, damage, etc.), to learn the relationship between them. Data sources (for storms as well as buoy data) are from NOAA and can be found in [1] and [2]. For our analysis, we focus on the Florida region. We picked buoys in the South, Southwest Florida, and a few buoys from the Gulf of Mexico. some of the stations shown in the figure below as well as at this link [3].

This is the buoy station list which was used:
41006 42001 42002 ALRF1 BURL1 CSBF1 GDIL1 SJLF1 BNKF1 VAKF1 MDKF1 THRF1 SMKF1 VCAF1 LONF1 SREF1 LMRF1 MUKF1 RKXF1 NPSF1 FMRF1 PMAF1 OPTF1 42013 42098 ARPF1 CDRF1 KTNF1 SHPF1 42036 TRDF1 PCLF1

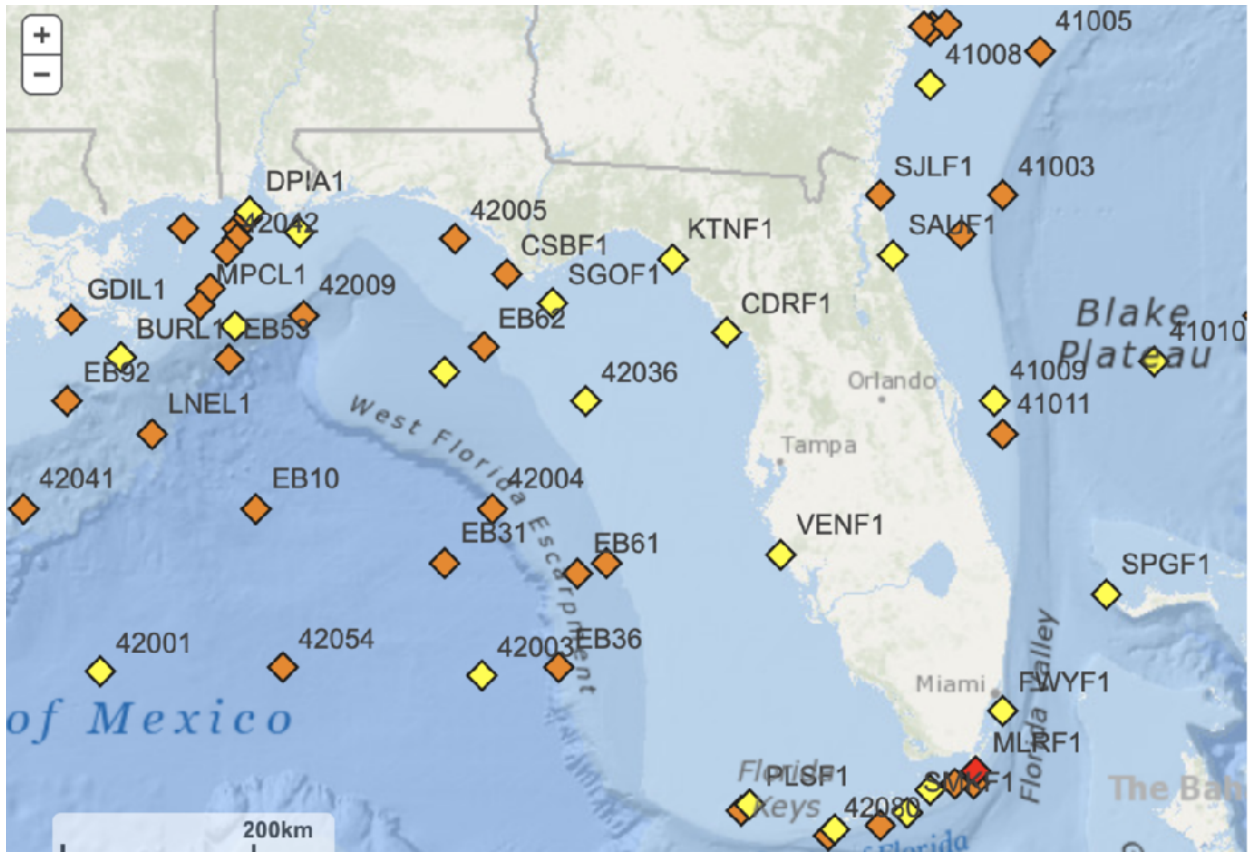

Figure: NOAA Buoy data in the Florida region (from the NOAA website)

There are a number of tar files by station and buoy and year which were downloaded, cleaned , aggregated (at the day level) and merged into a data frame in R. This was further merged with the event data. The time series data used is from 1985 to present.
These are the fields present in the buoys/stations dataset--
Physical variables measured and aggregated to hourly timestamps: Wind temperature, direction, tide levels, air temperature, sea surface temperature, pressure and so on.
These are the fields present in the extreme event/storm dataset--

At a daily aggregate: storm location, magnitude (high winds mainly), damage level, injuries, free text narrative comments and other fields [5]

**Missing data imputation for multivariate time series**

About 12% of the data is missing, sometimes for certain isolated time stamps (hourly) at random, and sometimes for larger segments (several hours in a day, or for full days). Since our forecasts are at a day level, we first aggregate data to a day level (using day means), and treat days with no valid values as missing. Then, we apply both univariate Kalman filter [9] and multivariate splines [10] for all variables, using data from nearby stations, for the same variable. E.g., for the wind speed time series, if values are missing for buoy 'A', and it has neighboring buoys 'B' and 'C', then the data across all these buoys are used to fill in missing values from buoy 'A' (spatio-temporal interpolation using multivariate splines for data imputation).

Below, we present imputation results, for univariate Kalman filter as well as for multivariate splines. The input with missing data is the variable air temperature across all neighboring buoys. Visually, e.g. buoys 42001 and 42002, it can be seen that multivariate splines do a better job than the univariate Kalman filter at reconstructing the time series missing values, because they leverage the correlations across the buoy recordings. We get the same result quantitatively as well (by testing both methods on 'fake' missing data created by inserting missing values, imputing the time series, and comparing with the original values).

Thus, we use multivariate splines for the data imputation step.

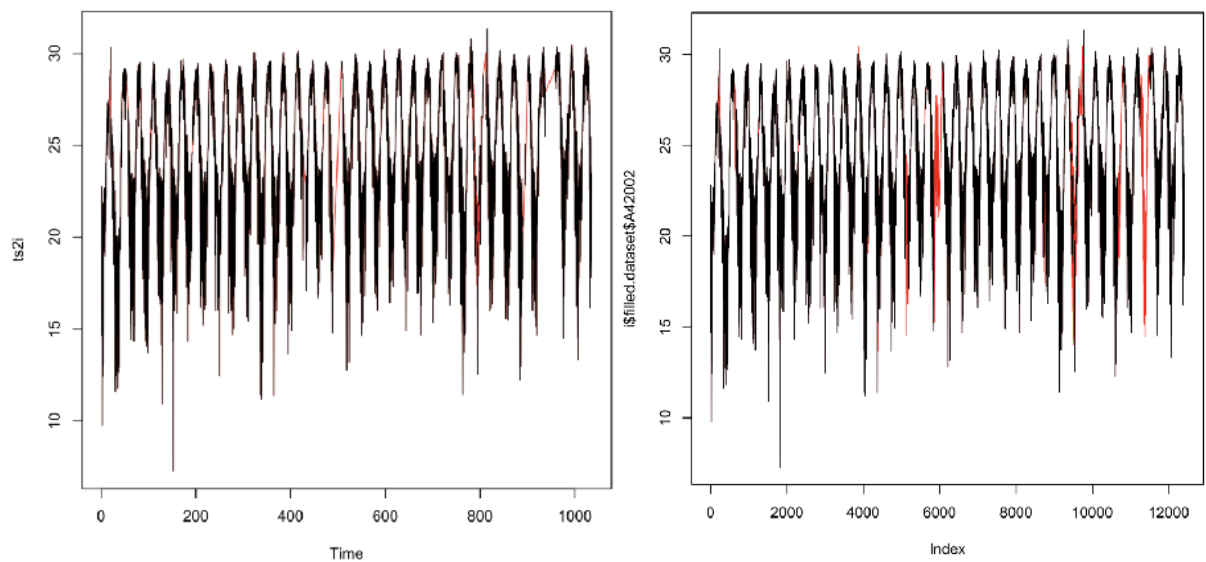

Figure: Missing data imputation using Kalman filter (left) versus multivariate splines (right): for buoy station 42002 air temperature time series. Black: original with missing data. Red: Imputed

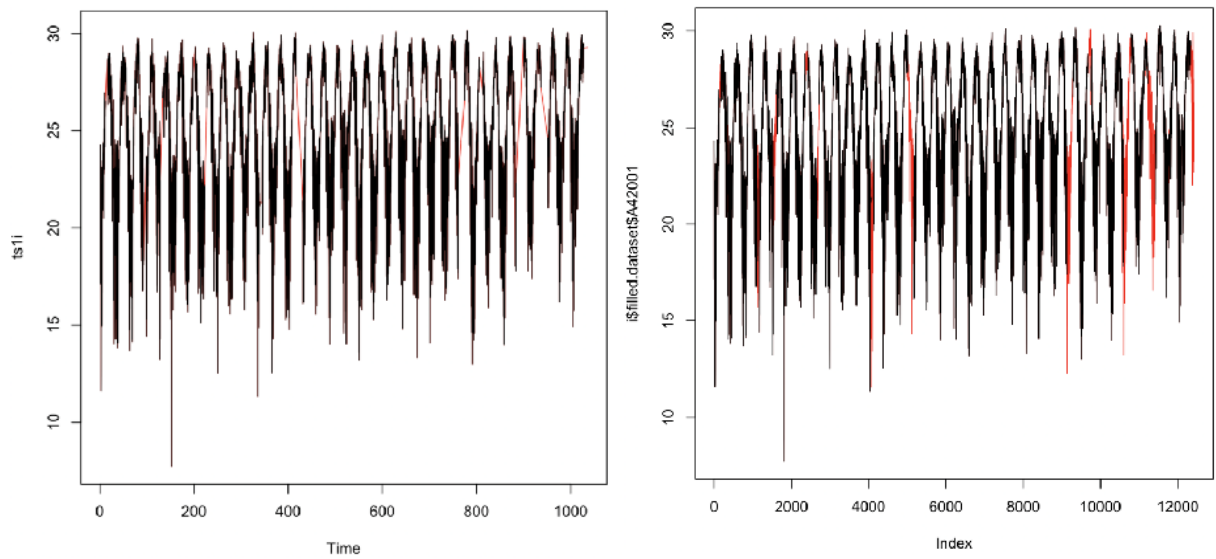

Figure: Missing data imputation using Kalman filter (left) versus multivariate splines (right): for buoy station 42001 air temperature time series. Black: original with missing data. Red: Imputed

## Statistical testing for trends of weather events

We decompose the extreme event magnitude time series using STL [11]. Then, we look at the trend of the storm magnitude time series, and test whether it is increasing over time, and whether the magnitude increase is statistically significant. Visually, the trend seems to be increasing, we verify this by running the augmented Dickey-Fuller test to check for stationarity, which returns significant p-values (>0.01), in this case, for accepting the null hypothesis that the series is non-stationary.

We also run the Ljung-Box test for independence, and it returns a p-value < 2.2e-16, which accepts the alternative hypothesis of non-independence for lagged values, thus suggesting a significant trend. Thus, there is an increasing trend for the magnitudes of extreme wind related weather events over time in the Florida region.

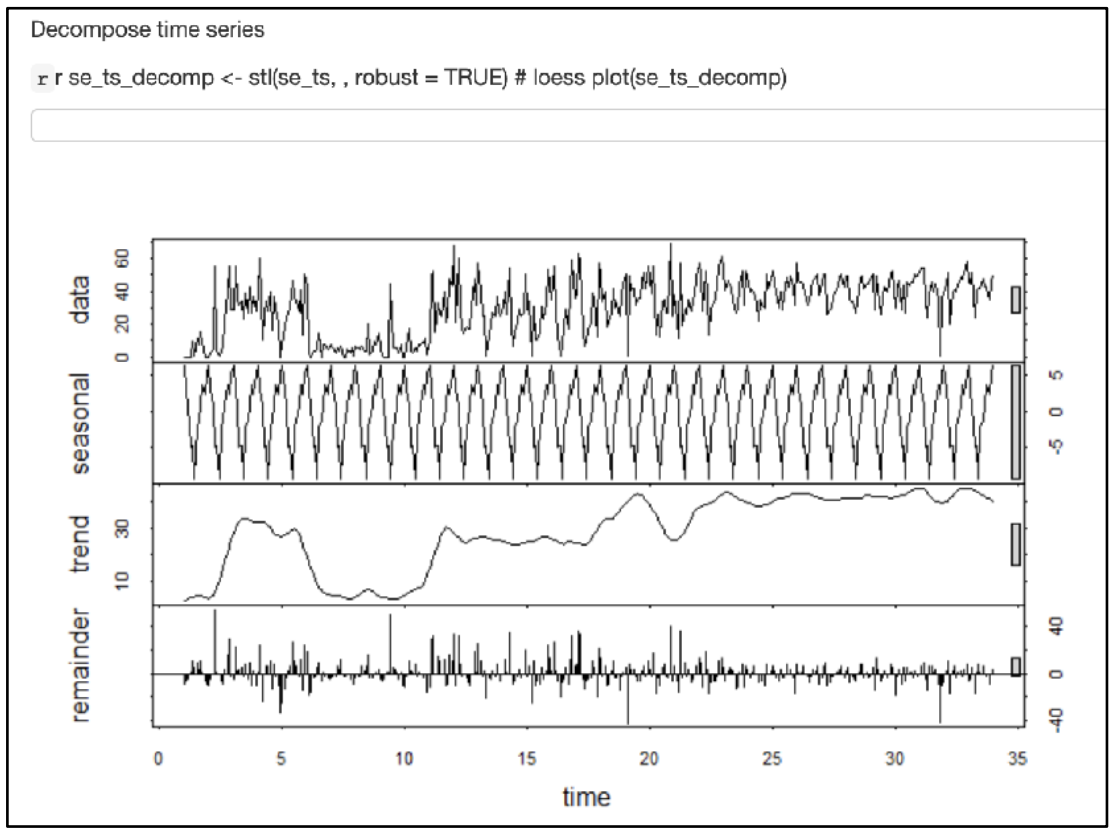

Figure: STL decomposition shows increasing trend of storm intensity over the years, which we confirmed using statistical tests (see text above for details)

**Correlation analysis of buoy variables with event time series and VAR forecast**

Once the data imputation is done, we fit linear and non-linear models using the lagged values of buoy variables to predict/forecast future values of storm intensity. Below, we illustrate this with a vector autoregressive model (VAR), which implicitly uses wave height (WVHT) lags/history as well as storm magnitude history/lags as the independent variables, and present values of storm magnitude (intensity) as well as present values of wave height as the dependent variable (across different time windows of fixed lengths).

The time domain, as well as cross-correlation plots for the two time series (wave height and storm magnitude) are shown below.

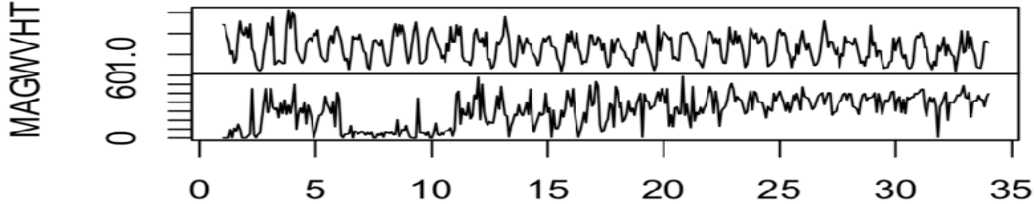

Figure: Values versus time-- Top: Wave height, bottom: Storm magnitude

```
ccf(wvht_mag_ts[,1], wvht_mag_ts[,2], type = "correlation")
```

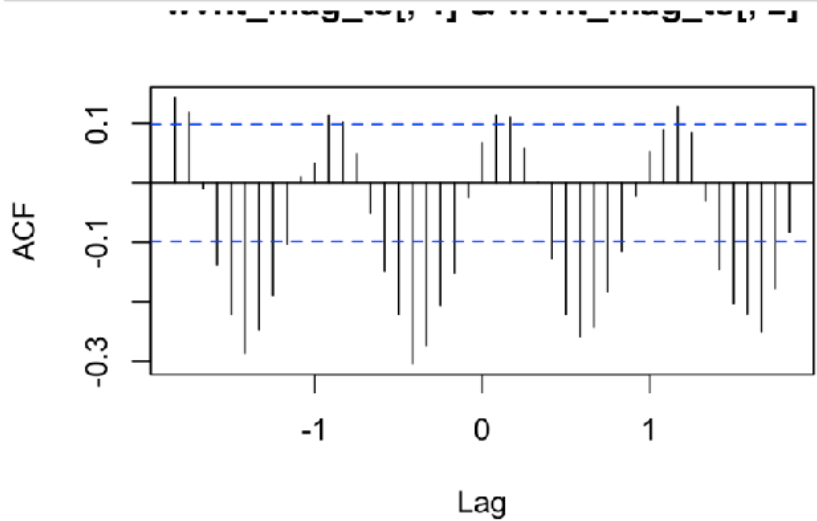

Figure: lags versus correlation between the two time series discussed in the text (storm magnitude and wave height)

The output of the VAR model is shown below, and lagged values of the wave height (WVHT--lag1, WVHT--lag6) and as well as event magnitude lags (MAG--lag1, MAG--lag2, trend) are deemed very significant variables (p value < 0.001) in the VAR predictive/forecast model, and the adjusted R-squared value is 0.8705, which shows that the VAR model is a good fit. AIC is used for model selection among the VAR models. Plots below show the fit versus actual values, as well as forecast time series for both the storm magnitude as well as the wave height.

Figure: VAR model output

```
wvht_mag_var <- VAR(wvht_mag_ts, type = "trend", lag.max = 6, ic = "AIC")
summary(wvht_mag_var$varresult$MAG)

Call:
lm(formula = y ~ -1 + ., data = datamat)

Residuals:
    Min      1Q  Median      3Q     Max
-40.887  -6.241   0.203   6.803  42.839

Coefficients:
        Estimate Std. Error t value Pr(>|t|)
WVHT.l1  6.903822   2.544600   2.713 0.006970 **
MAG.l1   0.344683   0.051508   6.692 7.97e-11 ***
WVHT.l2 -5.820210   3.257969  -1.786 0.074827 .
MAG.l2   0.158833   0.054471   2.916 0.003758 **
WVHT.l3  2.265851   3.274332   0.692 0.489350
MAG.l3   0.103167   0.055095   1.873 0.061905 .
WVHT.l4 -2.669935   3.275324  -0.815 0.415490
MAG.l4  -0.014117   0.054978  -0.257 0.797493
WVHT.l5 -6.257005   3.274008  -1.911 0.056746 .
MAG.l5   0.027987   0.054162   0.517 0.605650
WVHT.l6  8.494280   2.464352   3.447 0.000631 ***
MAG.l6   0.032690   0.052436   0.623 0.533377
trend    0.034382   0.008147   4.220 3.06e-05 ***
---
Signif. codes:  0 '***' 0.001 '**' 0.01 '*' 0.05 '.' 0.1 ' ' 1

Residual standard error: 12.54 on 378 degrees of freedom
Multiple R-squared: 0.8748,    Adjusted R-squared: 0.8705
F-statistic: 203.1 on 13 and 378 DF,  p-value: < 2.2e-16
```

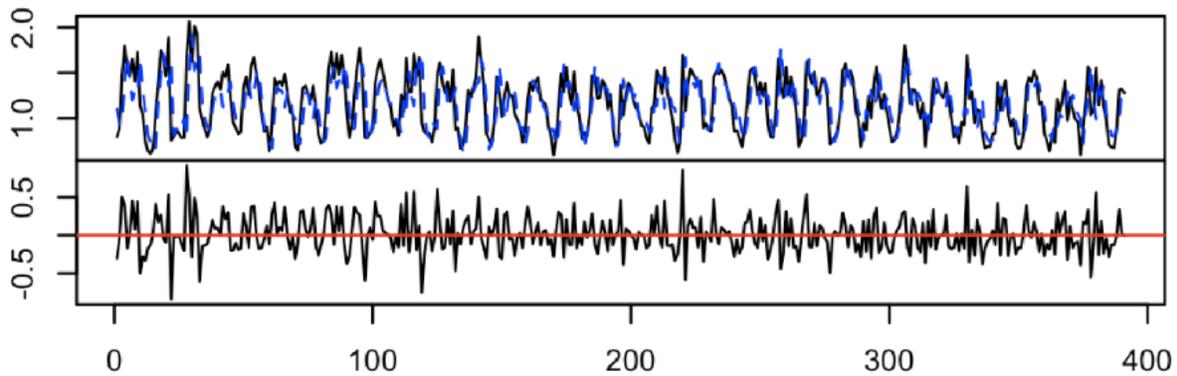

Figure: The wave height fit using the VAR model (blue) overlaid on actual values, and the residual in the below plot.

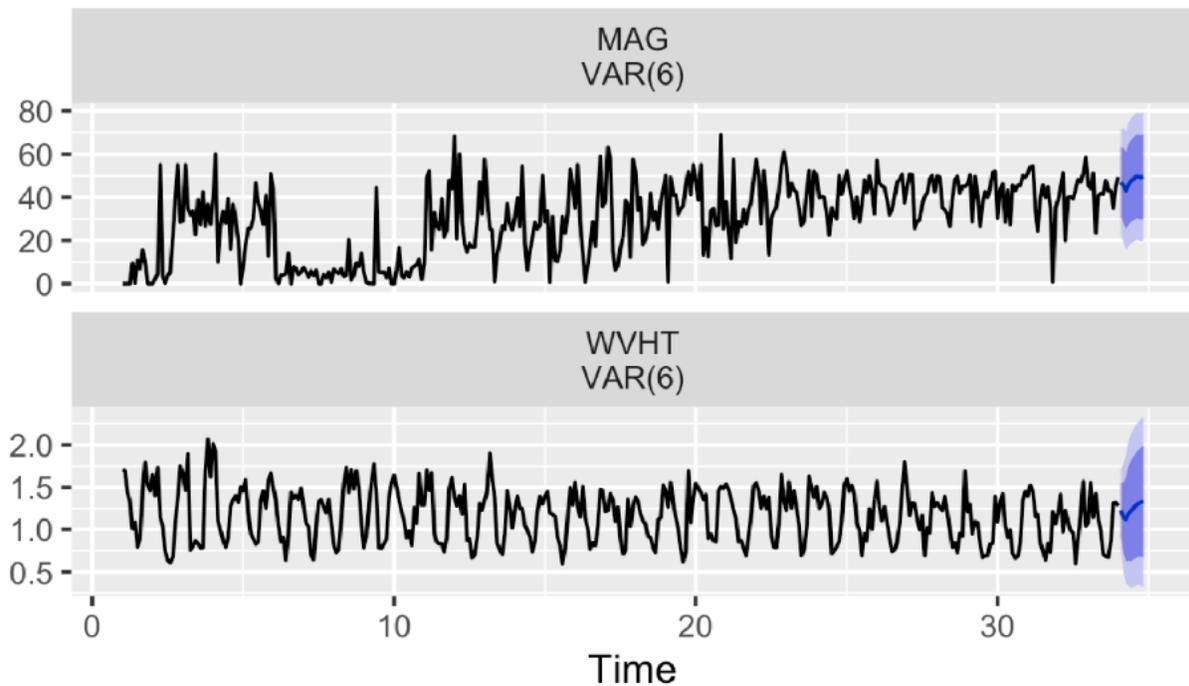

Figure: Forecast for both storm magnitude (top) and wave height using the VAR model

**Machine learning with buoy variables to predict event severity**

Extending the previous section beyond traditional time series forecasting models (such as VAR), machine learning based models are fit. To illustrate this concept, an Xgboost [12] model is fit to the time series data (this is done by dividing the time series into different time windows, and using the values in each window as the independent variable vector to predict the next value as the dependent variable, via the machine learning model), to predict the storm event levels (a regression problem). The model was fit on historic data (20 years) and used to predict the most recent 15 years. A root mean squared error of 11.2 was obtained on the train set, and 14.6 on the test set The results of the fit model and variable importance plot is below. Additionally, we also convert this to a binary classification problem, where we are trying to predict whether an event will be high intensity or not. This is the confusion matrix obtained for the binary classifier.

|  | Ground truth storm intensity (high or low) | |
|---|---|---|
| Predicted storm intensity | Low | High |
| Low | 2657 | 137 |
| High | 696 | 236 |

Table: Confusion matrix for a binary classifier (to predict high/low storm intensity)

As can be calculated from the above table, the true positive rate for 'normal' (i.e. low intensity) storm events is 77 % , and the true positive rate for the high intensity/extreme events class prediction is 64%.

Variable importance is plotted below.

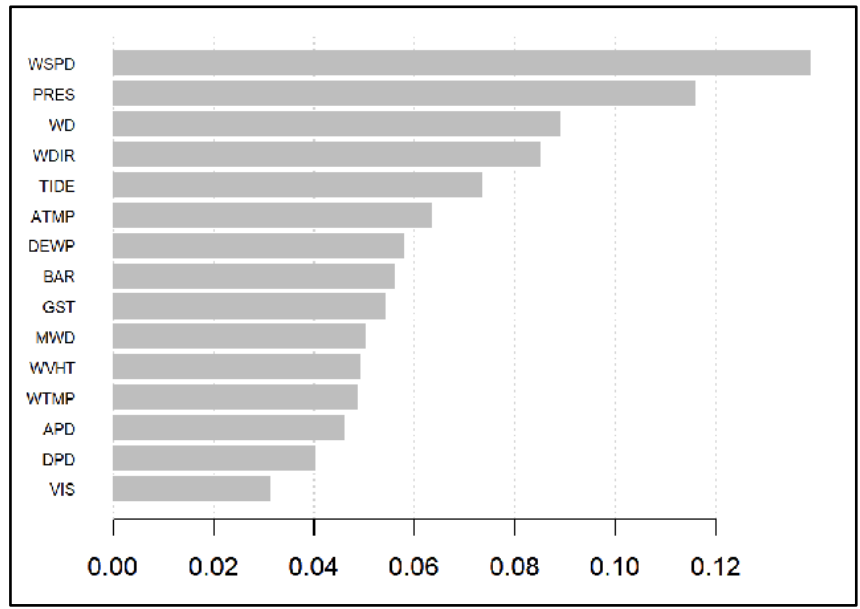

Figure: Variable importance plots for the Xgboost model, to forecast storm event intensity using buoy variables (Top variables are the most recent recorded values of the wind speed, pressure, direction, tide levels and atmospheric temperature)

## Conclusion

In this paper, we have proposed using machine learning to predict extreme weather events (storm magnitude) using buoy variable time series data. We have used multivariate splines for missing data imputation, as well as VAR and Xgboost for spatio-temporal forecasting.